# Low-loss Waveguide Fabrication by Inkjet Printing on Porous Silicon

**Alexander Kellarev, Shlomo Ruschin**
*Department of Physical Electronics, Faculty of Engineering,
Tel-Aviv University, Tel-Aviv 6997801, Israel*
kellarev@post.tau.ac.il; ruschin@eng.tau.ac.il

## Abstract

We report fabrication of optical waveguides embedded in oxidised porous silicon in a mask-less process. Patterned pore-filling was attained by means of inkjet printing, enabling lateral modification of the waveguide's index of refraction and other optical properties thus providing full two-dimensional refractive index control of the optical waveguide. The waveguides were created by applying a KDP salt solution over a porous silicon film by means of a material printer. Light guiding was demonstrated at wavelengths 1.064 and 1.550 µm. Measured values of the propagation loss at these wavelengths were 2.5 dB/cm and below 1.3 dB/cm respectively.

## Introduction

Optical waveguides are essential elements for a variety of applications in optical and opto-electronic systems such as optical communication and sensing, both as single elements and as parts of optical integrated circuits. Currently the creation of integrated optical elements on solid substrates is performed using relatively complex and costly methods, such as thin film deposition, ion implantation, thermal diffusion and laser-induced material modification often requiring multiple lithography masks and/or growing of additional structure layers. [1] These methods usually require complex processing equipment, e.g. vacuum systems, ion sources and powerful lasers.

Use of porous silicon for creation of channel waveguides has been reported previously by several research groups. [2–7] However, the processes used for the waveguide fabrication were also rather complex and involved fabrication steps like laser oxidation, photolithography, focused proton beam or growth of additional masked layers.

To address the complexity of the above mentioned methods, we propose and demonstrate a simple and effective technique of creation of waveguides and other planar optical elements by patterned pore filling in a porous thin film (patent pending). Pore filling results in increased index of refraction of the substrate and allows the lateral refractive index modification of the patterned regions. In a recent article we reported the modification of optical properties of oxidized porous silicon film by pore filling. [8] This process is relatively simple and able to create regions with desired index of refraction when applied with a mask or other patterning techniques. In contrast to the other methods used for waveguide fabrication in porous silicon, pore filling possesses the flexibility of tuning the refractive index of the waveguide core by varying the concentration of the filling material. In this letter we report the extension of this method by implementing it by inkjet printing. The technology of inkjet printing is nowadays emerging in the fields of micro-electronics and micro-optics [9,10], acquiring interest also for the fabrication of optical waveguides as reported very recently. [11–14] Embedded waveguides as those disclosed here appear advantageous regarding compatibility with current silicon-based processing and for coupling with customary silica fibres. Pore filling by salt solution enables lateral control of the waveguide's refractive index and other optical properties, and since the refractive index in porous silica films can be controlled in the depth direction by varying the current density during the electro-chemical processing, our technique enables the full 2D profiling of optical waveguides.

Light propagation through the created waveguides was examined at wavelengths 1.064 and 1.550 μm. The transmitted light was measured by means of end-facet imaging and fibre coupling. Attenuation coefficients were mainly obtained from the analysis of the light scattered from above the surface. For 1.55-μm wavelength radiation propagation loss coefficients below 1.3 dB/cm were measured. Total fibre-to-fibre insertion loss of the waveguides was also measured using a broadband source in spectral range 1.1..1.75 μm. In the following we present a short theoretical introduction and describe the main processing steps and methods used for characterisation of the waveguides.

## Theoretical background

Porous silicon (PSi) is a metamaterial, which was discovered by A. Uhlir in 1956. [15] The remarkable properties of this material are extremely large surface area and high ability to absorb other substances in its pores, what makes it an attractive candidate for using in various fields ranging from optoelectronics to biomedical applications.

PSi is usually created by means of electrochemical etching of Si wafers, which results in formation of a porous layer on top of the wafer. [16] The thickness of the layer and volume fraction of the pores (porosity) are determined by the Si resistivity and the etching current density and time. It is possible to create several porous layers with different porosities on the same substrate by etching with different current densities, enabling control of the refractive index in the depth dimension. When the pore diameter is small compared to the wavelength of light, PSi can be considered as an optical metamaterial. The effective index of refraction of porous silicon can then be well described by the generalised Bruggeman model, which represents the effective susceptibility of a composite medium by means of susceptibilities and volume fractions of the composites and depolarisation factors. [17,18] According to this model, replacing the voids in porous material will lead to altering its optical properties. [8] Specifically, insertion of any material with refractive index greater than one into the pores will result in an increase of the refractive index of the porous layer. Following this principle, a localised modification of the refractive index is possible by applying a patterned pore filling to a porous substrate, what allows creation of optical waveguides and other light guiding structures.

Pore filling of oxidised porous silicon (OPSi) with salts was recently reported by us. [8] Application of an aqueous salt solution on OPSi followed by drying results in the increase of the refractive index of the OPSi, the amount of which can be controlled by the solution concentration. Particularly, KDP salt ($KH_2PO_4$) is suitable for pore filling because it has good solubility in water and it is transparent in the visible region.

Patterned pore filling applied to double-layer OPSi creates regions of rectangular shape in the top layer, where the filling material remains. Such region can be observed in Figure 1, which shows a cross-section of a test sample created with saturated salt solution in order to increase the contrast of the refractive index. As can be seen in the image, the region is laterally well confined, indicating that the pores are not interconnected. The solution is mainly held in the top layer, where the capillary effect is stronger due to the smaller pore diameter.

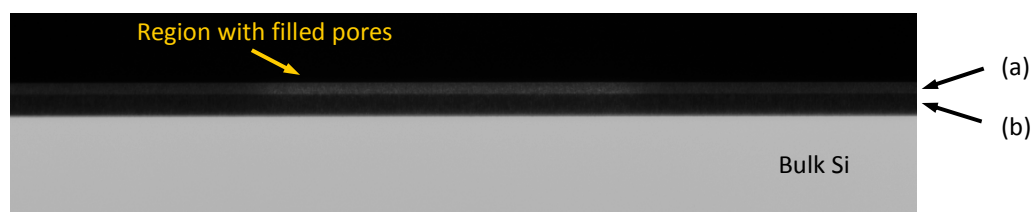

Figure 1: microscope image of cross section of the double-layer OPSi with filled pores; (a) – low porosity layer, (b) – high porosity layer.

## Fabrication

As a substrate for pore filling we fabricated samples with double-layer oxidised porous silicon. The samples were created by electrochemical etching of a (100)-cut p-type Si wafer with resistivity of 0.02..0.03 Ohm·cm. Two layers of different porosity were obtained by etching at different current densities.

The etching took place in a mixture of HF (48%) with $C_2H_5OH$ in proportion 1:1 by volume. After the etching the samples were thermally oxidised by means of a two-step process: pre-oxidation at 300⁰C followed by oxidation at 900⁰C for 2 hours. The oxidation was necessary in order to environmentally stabilise the samples and to ensure their hydrophilicity. The bottom layer possesses higher porosity and lower index of refraction than the top layer and acts as a cladding. The difference in the indices of refraction of the two layers was 0.15, measured as described in [8]. The etching parameters and the resulting properties of the porous layers in each of the samples are summarised in Table 1. The layer thickness was measured from microscope images of the cross-section of the samples. Indices of refraction of the porous layers were inferred from reflection measurements of single-layer samples etched at the same current densities.

Table 1: fabrication and optical parameters of the porous layers.

| Layer | Current density (mA/cm$^2$) | Etching time (s) | Thickness (μm) | Refractive index at 1.064 μm | Refractive index at 1.550 μm |
|---|---|---|---|---|---|
| top | 30 | 140 | 3.3 | 1.352 | 1.338 |
| bottom | 50 | 200 | 6.6 | 1.204 | 1.190 |

For the pore filling we used unsaturated aqueous solution of KDP ($KH_2PO_4$) salt. The dry salt was acquired from Sigma-Aldrich. The solution with concentration 5.1% (by weight) was prepared by dissolving dry salt in distilled water. After drying, the pore filling results in increase of 0.04 of refractive index of the top layer.

Several waveguides were printed with the KDP solution by means of a Super Inkjet material printer (SIJ Technology, Inc., Japan). Size of the cross-section of the waveguides was 21 μm x 3.3 μm, measured with an optical microscope. Sides of the samples were cleaved to allow access to the ends of the waveguides. After the cleaving, the lengths of the waveguides were 5 mm.

## Optical measurements

Several independent methods were implemented for the optical characterization of the waveguides. In all of them, laser sources were coupled into the waveguides through a lensed fibre. In order to acquire transversal mode profiles, the transmitted light was imaged at the opposite side through a 20x microscope lens. For the transmission measurements, the camera and the lens were replaced by a bare SM-28 fibre. Spectral transmission in range 1.1..1.75 μm was measured using a NKT Photonics SuperK Compact laser with VIS/IR splitter as a source and Ando AQ-6315B optical spectrum analyser as a detector. An additional camera with 5x microscope objective was located above the sample in order to collect scattered light from the sample's surface. The scattered light images allowed the measurement of the lateral intensity profile of the guided light and the propagation loss.

## Results and discussion

Light propagation was measured at wavelengths 1.064 and 1.550 µm. As expected, we found that the waveguides support multimode propagation at both wavelengths. Single-mode propagation was attained by proper alignment of the lensed fibre. Examples of images of the light transmitted through a waveguide are shown in Figure 2. Spatial profiles of the transmitted light along horizontal and vertical directions were fit to a Gaussian function in order to obtain the mode width at 1/e of the peak intensity. The measured mode widths were 7.4 µm x 2.9 µm for wavelength of 1.064 µm and 13.2 µm x 4.2 µm for wavelength of 1.550 µm. These results are comparable with theoretical mode widths of 12.5 µm x 2.1 µm and 12.7 µm x 2.2 µm respectively, which were calculated by means of Marcatili's method. [19]

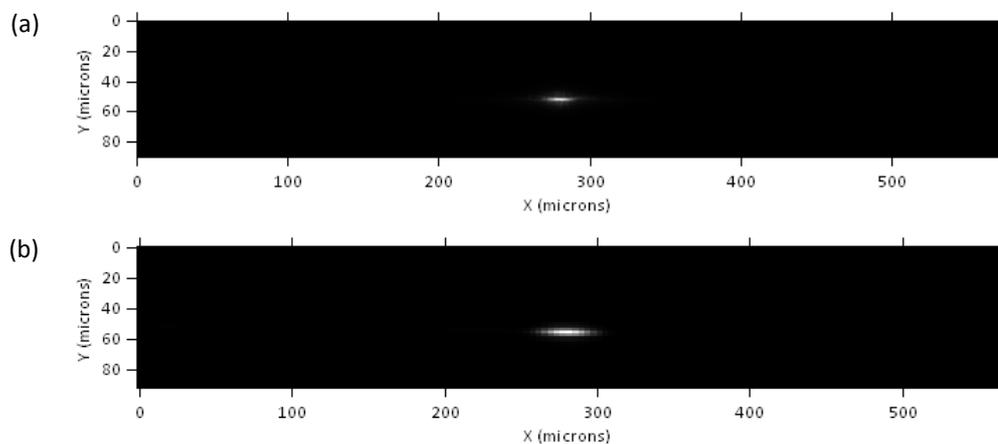

Figure 2: images of the single modes transmitted through the waveguides; (a) - 1.064 µm, (b) - 1.550 µm.

Typical images of light scattered from the samples' surface are presented in Figure 3. As shown, some amount of scattered light appears in regions adjacent to the waveguides. Appearance of this light is attributed to volume scattering in the porous layers and interface scattering at the waveguide bounds. In fact, the top porous layer supports propagation of scattered light, since its index of refraction is higher than that of the bottom layer and it acts as a planar waveguide.

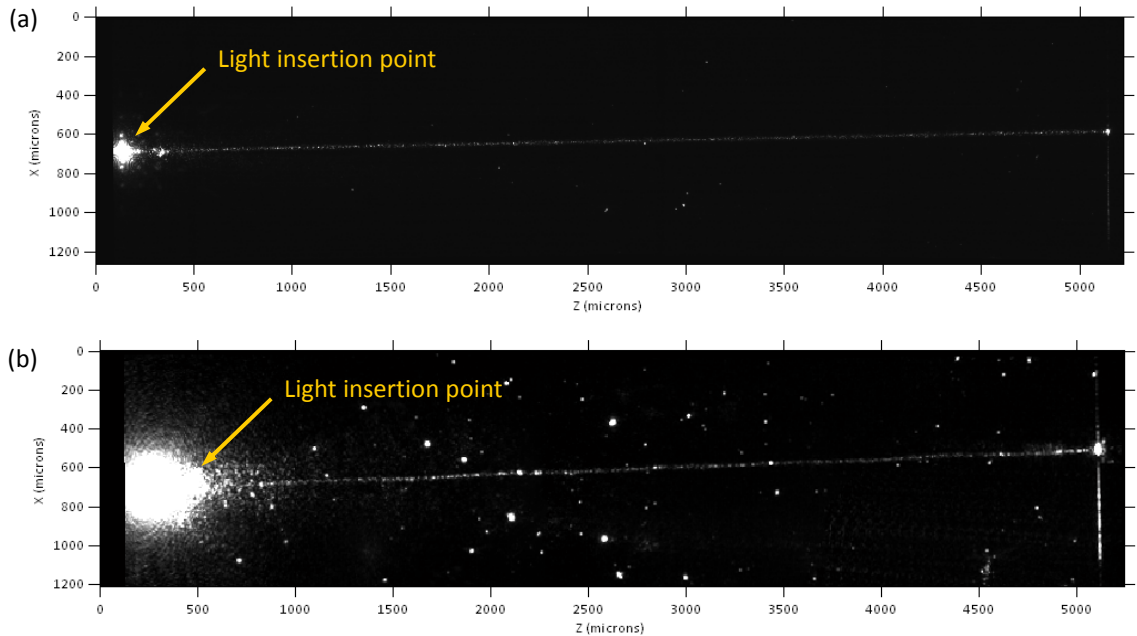

Figure 3: images of the light scattered from the top surface; (a) - 1.064 μm, (b) - 1.550 μm.

Propagation loss along the waveguides was found from fitting the intensity profiles of the scattered light to an exponential function. [2,5,7,20,21] The intensity profiles were extracted by designating a strip of an appropriate width along the waveguides. Examples of the extracted and the fitted profiles are shown in Figure 4. The calculated attenuation and loss values appear in Table 2. The obtained results show that losses at 1.064 microns are nearly twice than those at 1.550 μm.

Table 2: attenuation measured for two waveguide profiles at two wavelengths.

| Wavelength | Attenuation (μm$^{-1}$) | Loss (dB/cm) |
| --- | --- | --- |
| **1.064 μm** | 0.00006±0.00007 | 2.5±0.8 |
| **1.550 μm** | 0.00003±0.00004 | 1.3±3.1 |

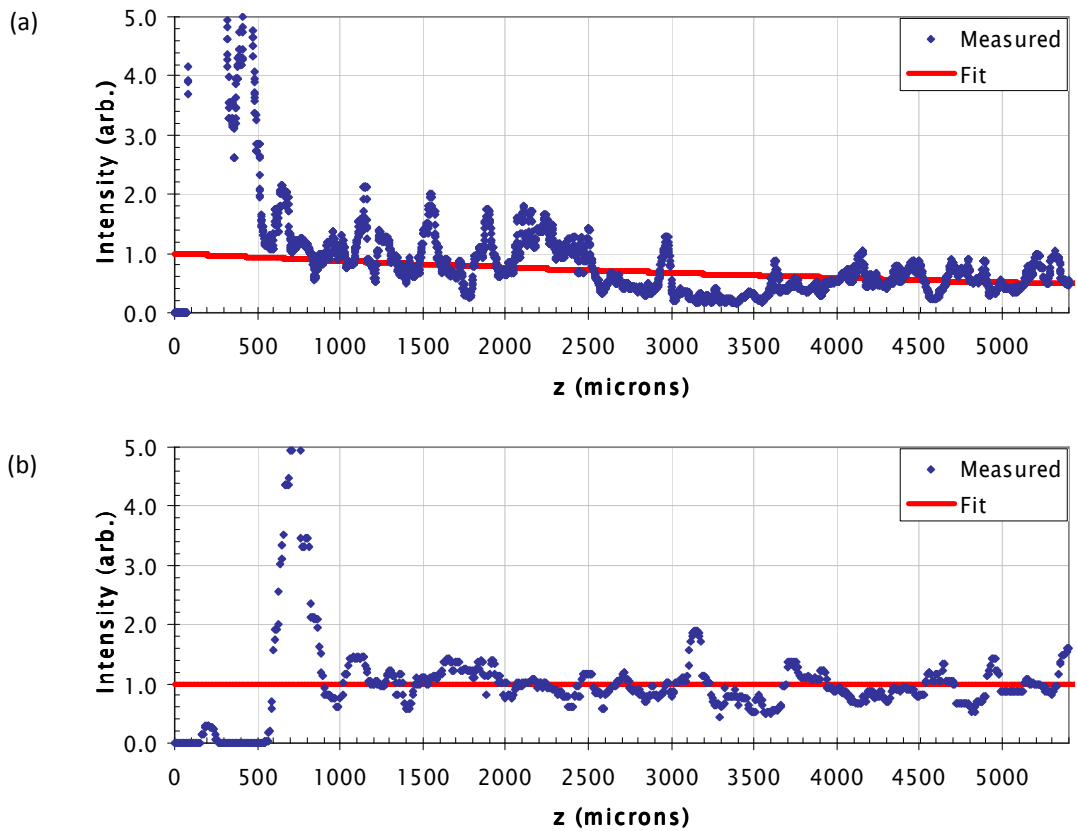

Figure 4: measured and fitted profiles of the light scattered along the waveguides; (a) - 1.064 μm, (b) - 1.550 μm.

In order to complement the measurements described above, we measured the insertion losses using a broadband source in conjunction with a spectrum analyser. Measurements of spectral transmission of three different waveguides in range 1.1..1.75 μm are presented in Figure 5. The transmission was determined taking as reference the transmission measured without the sample, when light was directly coupled between the input and output fibres. Fibre-to-fibre insertion losses of 3.8 dB and 4.6 dB were measured at wavelengths of 1.064 and 1.550 μm respectively. The lowest insertion loss value is 2.4 dB, attained at 1.13 μm. These values included also the unpolished edge losses which apparently are responsible for a major portion of the total insertion loss, as evidenced in Figure 3.

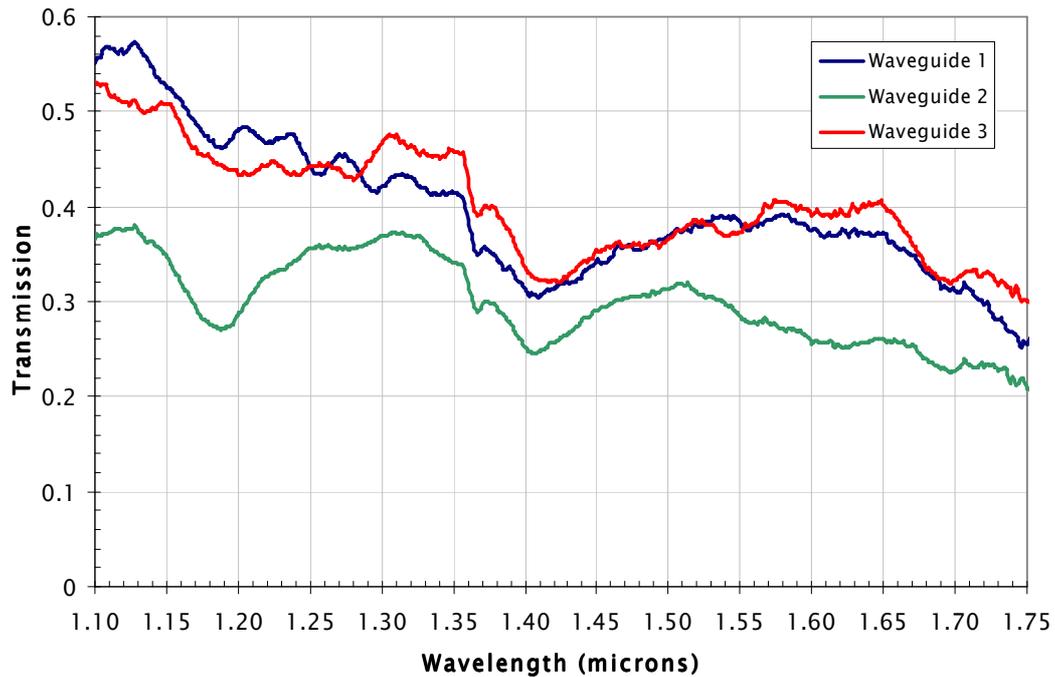

Figure 5: measured spectral transmission.

## Conclusion

We reported here the first demonstration of optical waveguides printed in porous silicon. We observed single-mode propagation at wavelengths 1.064 and 1.550 µm. The measured propagation losses were found to be relatively low and competitive with other reported values for patterned waveguides in PSi.

Fabrication of embedded waveguides by printing in a porous substrate appears as a novel and advantageous technique: it enables true two-dimensional profile control of refractive indices and it is compatible with current silicon-based microelectronics processing. This method may be further advanced by exploration of various filling materials and refinement of printing techniques.

## Acknowledgements

The authors are grateful to the specialists from SIJ Technology, Inc. for their help in this work.